\documentclass[final,11pt]{elsarticle}
\usepackage{amsmath}
\usepackage[margin=2.5cm]{geometry}
\usepackage{physics}
\usepackage{siunitx}
\usepackage{subcaption}

\usepackage[parfill]{parskip}

\usepackage{color}
\usepackage{graphicx}
\usepackage{grffile}

\usepackage[hidelinks]{hyperref}

\journal{Computers \& Geosciences}

\biboptions{sort&compress}

\begin{document}

\title{Modeling Transport of Charged Species in Pore Networks: Solution of the Nernst-Planck Equations Coupled with Fluid Flow and Charge Conservation Equations}

\author[UW]{Mehrez Agnaou\fnref{footnote1}}
\author[UW,MG]{Mohammad Amin Sadeghi\fnref{footnote2}}
\author[UCL]{Thomas George Tranter\fnref{footnote3}}
\author[UW,MG]{Jeff Gostick\corref{cor}\fnref{footnote4}}
\cortext[cor]{Corresponding author}
\ead{jgostick@uwaterloo.ca}

\address[UW]{Department of Chemical Engineering, University of Waterloo, Waterloo, ON, Canada}
\address[MG]{Department of Chemical Engineering, McGill University, Montreal, QC, Canada}
\address[UCL]{Department of Chemical Engineering, University College London, London, United Kingdom}

\fntext[footnote1]{Model development, code implementation, manuscript drafting.}
\fntext[footnote2]{Model development, code implementation.}
\fntext[footnote3]{Code implementation, revising the manuscript.}
\fntext[footnote4]{Study design, code implementation.}

\begin{frontmatter}
\begin{abstract}
A pore network modeling (PNM) framework for the simulation of transport of charged species, such as ions, in porous media is presented. It includes the Nernst-Planck (NP) equations for each charged species in the electrolytic solution in addition to a charge conservation equation which relates the species concentration to each other. Moreover, momentum and mass conservation equations are adopted and there solution allows for the calculation of the advective contribution to the transport in the NP equations.

The proposed framework is developed by first deriving the numerical model equations (NMEs) corresponding to the partial differential equations (PDEs) based on several different time and space discretization schemes, which are compared to assess solutions accuracy. The derivation also considers various charge conservation scenarios, which also have pros and cons in terms of speed and accuracy. Ion transport problems in arbitrary pore networks were considered and solved using both PNM and finite element method (FEM) solvers. Comparisons showed an average deviation, in terms of ions concentration, between PNM and FEM below $5\%$ with the PNM simulations being over ${10}^{4}$ times faster than the FEM ones for a medium including about ${10}^{4}$ pores. The improved accuracy is achieved by utilizing more accurate discretization schemes for both the advective and migrative terms, adopted from the CFD literature. The NMEs were implemented within the open-source package \texttt{OpenPNM} based on the iterative Gummel algorithm with relaxation.

This work presents a comprehensive approach to modeling charged species transport suitable for a wide range of applications from electrochemical devices to nanoparticle movement in the subsurface. \\
\end{abstract}

\begin{keyword}
Porous media \sep Nernst-Planck equations \sep pore network modeling \sep \texttt{OpenPNM}
\end{keyword}
\end{frontmatter}



\section{Introduction}\label{sec:introduction}

The Nernst-Planck equations are widely used in the literature to describe the transport of ionic species in electrochemical systems \cite{meng2014,metti2016}. With respect to porous media, the equations describe ion transport in a wide variety of applications such as electrochemical cells \cite{vansoestbergen2010} and certain redox flow batteries \cite{sadeghi2019b}. They are also used to analyze ion conduction in biological structures of pores \cite{bolintineanu2009}, but probably the most common applications are for the study of ion transport mechanisms in clay soils and concrete. \citet{smith2004} applied the NP equations to the analysis of transport through platy-clay soils and \citet{pivonka2004} analyzed chloride diffusion in concrete for the estimation of structural degradation due to corrosion. Moreover, it has been shown that simulations based on the NP equations accurately predict ionic diffusion coefficients experimentally estimated on concrete \cite{narsilio2007}. In a more recent work \cite{azad2016}, the transport processes in a system including a concrete plug surrounded by clay stone were modeled using the NP equations.

Another important field where the NP equations are used is modeling transport in capacitive charging and deionization \cite{biesheuvel2010, gabitto2015}. Comparisons between simulation results and experimental data \cite{sharma2015} highlighted the capabilities of the NP based simulations to help in the design of capacitive deionization devices. While the transport of ionic species in the bulk of a solution flowing through a porous medium is generally described using the NP equations, a charge conservation equation is required to close the system. One option, perhaps the most accurate, uses the well-known Poisson equation for the electrostatic potential \cite{newman2012}. The Poisson equation relates the electric charge density to the Laplacian of the potential and describes the movement of the charged species in solution. This yields the famous Poisson Nernst-Planck system of equations. Charge conservation can also be enforced through a Laplace equation for the potential which allows for further mathematical simplifications under certain assumptions \cite{newman2012}. In the presence of fluid flow, the solution of the flow problem based on the mass and momentum conservation equations (Stokes or Navier-Stokes) enables the calculation of the advective term in the NP equations.

Solving electrochemical problems in porous media at the pore-scale based on the NP equations is generally carried-out using computational mesh that conforms to the real geometry of the system being analyzed. Different methods have been used to numerically solve the transport equations such as the finite difference \cite{bolintineanu2009,meng2014,sharma2015} and finite element \cite{samson1999,narsilio2007,lu2010,metti2016,azad2016}. However, it is well-known that direct numerical simulations (DNS) require significant computational resources. The same logic applies to many other transport problems such as pure diffusion or dispersion in porous media. PNM, as an alternative pore-scale modeling approach, requires substantially lower computing resources (compared to pore-scale DNS) and have been successfully applied to study physics such as diffusion reaction \cite{gostick2007} and dispersion \cite{sadeghi2019a} in porous media. However, the use of PNM to study transport of charged species is in its infancy. For instance, in a study of electrokinetic transport through charged porous media \cite{obliger2014}, a steady-state PNM approach was used. This work \cite{obliger2014} is one of the first modeling electrochemical systems based on PNM. The used pore-scale microscopic transport coefficients were simple analytical relations obtained by solving the NP equations in a cylinder. Recently \cite{lombardo2019}, a pore network model based on the NP equations was used to study porous electrodes in electrochemical devices. However, their approach \cite{lombardo2019} was based on the upwind scheme, which was recently shown to have high errors when P{\'e}clet number is above unity \cite{sadeghi2019a}.

In this work, a more accurate method was developed and validated to solve the charge conservation NP system in pore networks. This new method will ultimately allow for accurate pore-scale simulation of transport in electrochemical systems with substantially lower computational cost compared to DNS approaches such as FEM. One aim of the present work is to identify the best approach among various options and to establish a numerically accurate and robust algorithm. Future work can then build on this solid foundation.

Although the simplifications related to PNM may induce additional errors into the numerical solution, it has been shown through comparisons between results of advection diffusion simulations, that the PNM approach provides reasonably accurate solutions \cite{yang2016} compared to those obtained from DNS using lattice Boltzmann and finite volume methods. This work presents a novel PNM framework for the simulation of charged species transport. The framework is based on highly accurate discretization schemes in addition to several charge conservation options. It also supports transient simulations and handles non-linear source terms.

\section{Background}\label{sec:background}

This work considers single-phase, isothermal, incompressible flow of a dilute electrolytic solution, treated as a Newtonian fluid, in a non-deformable porous medium. Assuming flow in the viscous-dominated regime \cite{agnaou2016,agnaou2017}, the movement of the electrolytic solution can be described using the following steady-state momentum and mass conservation (Stokes) equations
\begin{equation}
\mu \laplacian{\vb*{u}}-\grad{p}=\vb*{0},\label{eq:momentum}
\end{equation}
and
\begin{equation}
\div{\vb*{u}}=0,\label{eq:mass}
\end{equation}
where $\vb*{u}$ is the velocity of the solution, $p$ its pressure, and $\mu$ its dynamic viscosity and is considered to be constant. Using the NP equation, the flux of ionic species $n$ in the solution is given by \cite{newman2012,biesheuvel2010,sharma2015}
\begin{equation}
\vb*{N}^{n}=-D^{n}\grad{c^{n}}+\vb*{u}c^{n}-\frac{D^{n}z^{n}F}{RT}c^{n}\grad{\phi},\label{eq:np_flux}
\end{equation}
where $c^{n}$ is the ion concentration, $\phi$ is the electrostatic potential, $D^{n}$ is the diffusion coefficient of species $n$ and $z^{n}$ its valence, and $F$ is the Faraday constant. Eq. \ref{eq:np_flux} as written follows several authors \cite{newman2012,sharma2015} defining the mobility based on the Nernst-Einstein equation, $u_{mob}^{n}=D^{n}/\qty(RT)$, where $R$ is the universal gas constant and $T$ a constant absolute temperature. The flux as defined by Eq. \ref{eq:np_flux} consists of three terms, representing different transport mechanisms namely, molecular diffusion, bulk advection, and electrostatic migration. Moreover, a mass conservation equation is considered for each of the ionic species $n$ as follows
\begin{equation}
\pdv{c^{n}}{t}=-\div{\vb*{N}^{n}}.\label{eq:np_con}
\end{equation}

Substituting the flux from the Nernst-Planck equation (Eq. \ref{eq:np_flux}) into the conservation equation (Eq. \ref{eq:np_con}), yields an equation for each of the ionic species as follows
\begin{equation}
\pdv{c^{n}}{t}=-D^{n}\laplacian{c^{n}}+\vb*{u}\vdot\grad{c^{n}}-\frac{D^{n}z^{n}F}{RT}\div(c^{n}\grad{\phi}).\label{eq:np}
\end{equation}

The governing equations for fluid flow and concentration of species (Eqs. \ref{eq:momentum}, \ref{eq:mass}, \ref{eq:np}) are now defined. However, an additional equation is required to close the system of equations since the electrostatic potential is unknown. In this work, three different approaches were considered. Using the Gauss electrostatic theorem \cite{newman2012}, one could relate the distribution of ions in the electrolytic solution to the variation of the electric field through a Poisson equation as follows \cite{smith2004,samson1999}
\begin{equation}
\div{\qty(\varepsilon \varepsilon_{r} \grad{\phi})}=-F\sum_{n}\qty(z^{n}c^{n}),\label{eq:poisson}
\end{equation}
such that $\varepsilon$ is the vacuum permittivity and $\varepsilon_{r}$ is the relative permittivity of the electrolytic solution. The quantity on the right-hand side (rhs) of equation \ref{eq:poisson} is the electric charge density per unit volume. The solution of the Poisson equation is numerically challenging due to numerical instabilities \cite{jerome1996,metti2016} and stabilization techniques are often required \cite{meng2014}. More stable and simpler alternatives to Eq. \ref{eq:poisson} can be used to close the system and enforce charge conservation. However, these alternative equations, discussed in what follows are derived based on specific assumptions and hence, their validity should be limited to specific cases \cite{macgillivray1968,macgillivray1969}. In fact, charge conservation can be imposed as follows
\begin{equation}
\div{\vb*{i}}=0,\label{eq:charge_conservation_01}
\end{equation}
where $\vb*{i}$ is the current density and is given by
\begin{equation}
\vb*{i}=F\sum_{n}\qty(z^{n}\vb*{N}^{n}).\label{eq:current_density_01}
\end{equation}
Replacing the flux $\vb*{N}^{n}$ in Eq. \ref{eq:current_density_01} by its value from Eq. \ref{eq:np_flux} yields
\begin{equation}
\vb*{i}=-F\sum_{n}\qty(z^{n}D^{n}\grad{c^{n}})+F\vb*{u}\sum_{n}\qty(z^{n}c^{n})-\frac{F^{2}}{RT}\grad{\phi}\sum_{n}\qty({z^{n}}^{2}D^{n}c^{n}).\label{eq:current_density_02}
\end{equation}
Then, by virtue of electroneutrality, $\sum_{n}{z^{n}c^{n}}=0$, the second term on the rhs of Eq. \ref{eq:current_density_02} is zero.
Insertion of Eq. \ref{eq:current_density_02} into Eq. \ref{eq:charge_conservation_01} gives
\begin{equation}
\div{\qty(K \grad{\phi})}=-F\sum_{n}\qty[z^{n}\div{\qty(D^{n}\grad{c^{n}})}],\label{eq:charge_conservation_02}
\end{equation}
where $K$ is the conductivity of the electrolytic solution and is given by
\begin{equation}
K=\frac{{F}^{2}}{RT}\sum_{n}\qty(z^{n\,2}D^{n}{c}^{n}),\label{eq:kappa}
\end{equation}
Finally, for negligible concentration gradients and assuming a uniform conductivity $K$, Eq. \ref{eq:charge_conservation_02} reduces to a Laplace equation for the potential as follows
\begin{equation}
\laplacian{\phi}=0.\label{eq:laplace}
\end{equation}

\section{Pore Network Modeling Formulation}\label{sec:pnm}

\begin{figure}
	\centering
	\includegraphics[width=0.6\linewidth]{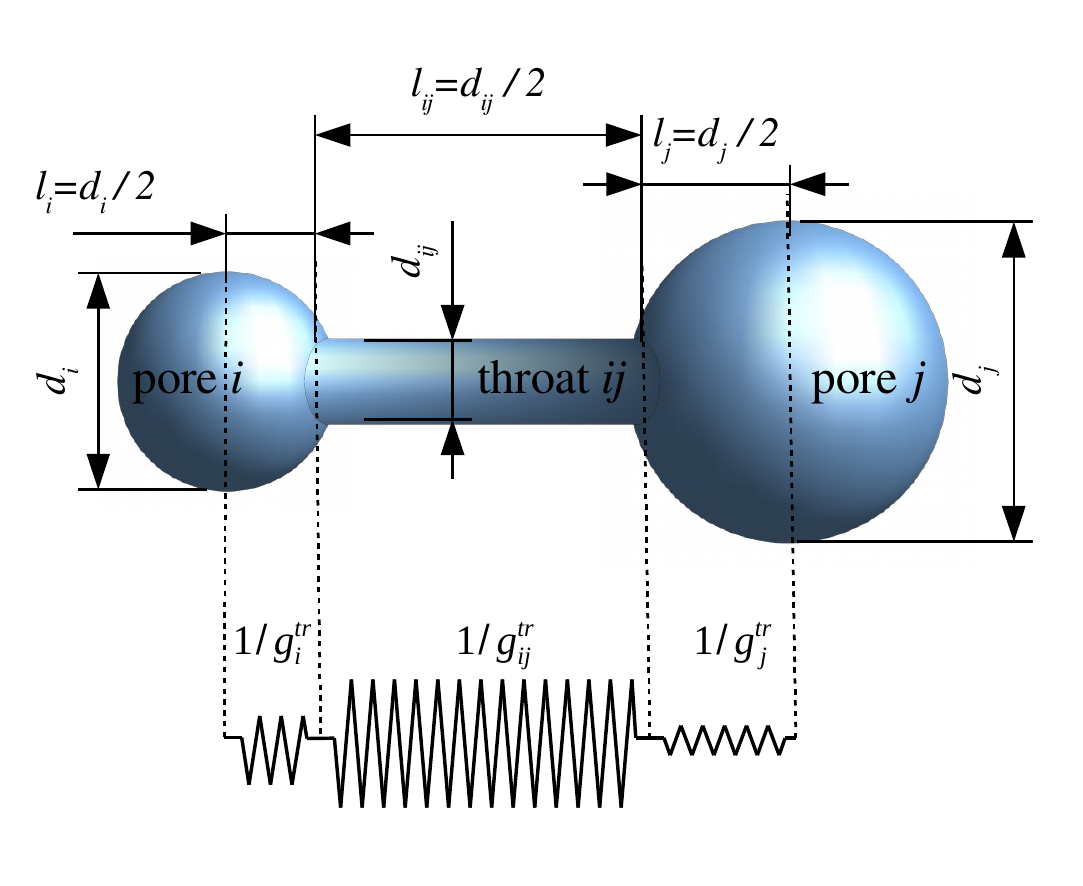}
	\caption{\label{fig:conduit} Pore-throat-pore assembly as a single conduit in PNM. Conduit made of throat $ij$ and halves of the neighbor pores $i$ and $j$ of diameters $d_{ij}$, $d_{i}$, and $d_{j}$ and lengths $l_{ij}$, $l_{i}$, and $l_{j}$ and opposing resistances to a transport mechanism $tr$ (from $i$ to $j$ and vice versa) of $1/g^{tr}_{ij}$, $1/g^{tr}_{i}$, and $1/g^{tr}_{j}$, respectively. Conductance of the assembly is given by Eq. \ref{eq:pnm_Gtr}.}
\end{figure}

The pore network is a simplified representation of a real porous medium geometry, consisting of pore bodies interconnected by throats. Figure \ref{fig:conduit} shows a pore-throat-pore conduit of a pore network. For the sake of simplicity regarding the conservation equations to be considered, idealized shapes are assigned to pores and throats. In this sense, and for a three-dimensional (3D) medium, pores and throats are generally represented by spheres and circular cylinders, respectively. For a two-dimensional (2D) geometry, pores and throats are described by circles and rectangles, respectively.

The conductance of the pore-throat-pore assembly or conduit for a given transport mechanism $tr$ (see Fig. \ref{fig:conduit}) is given, from the linear resistor theory for resistors in series \cite{gostick2007}, by
\begin{equation}
	G^{tr}_{ij}=\qty(\frac{1}{g^{tr}_i}+\frac{1}{g^{tr}_{ij}}+\frac{1}{g^{tr}_j})^{-1}.\label{eq:pnm_Gtr}
\end{equation}
Efficient algorithms for the extraction of pore networks from 2D and 3D images are available in the literature \cite{dong2009,rabbani2014,gostick2017}, even for dual networks \cite{khan2019}, within the open-source image analysis package \texttt{PoreSpy} \cite{gostick2019}.

This work assumes perfect mixing of the solute within the pore space, unlike more sophisticated approaches to be discussed below. In addition, conservation of physical quantities are enforced in the pores only. Therefore, for a time dependent transport problem, the total void volume of the porous medium is assigned to the pores whereas the throats are considered to have a zero volume. The volume of the throats is distributed among their neighboring pores. This approach offers simplicity and computational efficiency which allows for pore-scale simulations at relatively lower computational costs compared to DNS.

The assumption of perfect mixing is robust for transport problems involving pure diffusion. When additional transport mechanisms such as advection come into play, this assumption remains valid at low P{\'e}clet numbers (P{\'e}clet numbers smaller than unity) where the P{\'e}clet is the ratio of advective to diffusive contributions. The validity of the perfect mixing assumption was extended to pore-scale P{\'e}clet numbers up to $257$ by \citet{mehmani2015} and by \citet{yang2016} in disordered sphere packs and up to $10$ by \citet{sadeghi2019a} in cubic networks of random pore sizes. Thus, the mixed-cell method can be used for modeling transport phenomena in disordered porous structures where moderate deviations from pure diffusion exist. The structural disorder refers in the present work to the randomness in the pores and throats sizes and in the coordination number of pores. Deviations from pure diffusion considered in this study (see section \ref{sec:comparisons}) result from advective and migrative fluxes. For ordered porous structures, the mixed-cell method should be reserved only for diffusion dominated problems. Furthermore, the mixed-cell method ignores the impact of non-uniform velocity profiles in pores and throats on the transport of chemical species. In the same manner as for the perfect mixing assumption, uniform velocity profiles were found to have a negligible effect on transport in disordered media \cite{mehmani2015,yang2016}. Consequently, the PNM method is appropriate to perform pore-scale simulations of advection diffusion problems (and advection diffusion migration problems as will be shown below) in disordered porous media at low computational costs. An alternative to the mixed-cell method when high concentration gradients are expected within the pore space, although computationally more expensive, is the streamline splitting approach \cite{mehmani2014}.

\subsection{Stokes Flow}

Given steady-state Stokes flow (Eqs. \ref{eq:momentum} and \ref{eq:mass}) of a Newtonian fluid, corresponding to the electrolytic solution, the mass conservation equation for an arbitrary pore $i$, is
\begin{align}
\sum_{j=1}^{N_i}{G}_{ij}^h\qty(p_i-p_j)=0, & \quad i=1,2,\dots,N_p,\label{eq:pnm_flow}
\end{align}
where the subscripts $i$ and $j$ correspond to the considered pore and the neighboring ones, respectively, and $p_i$ and $p_j$ are the pressure values in pores $i$ and $j$, respectively. In Eq. \ref{eq:pnm_flow}, $N_i$ is the number of pores neighboring of pore $i$, $N_p$ is the total number of pores in the network, and $G_{ij}^h$ is the hydraulic conductance of the pore-throat-pore assembly and is given by Eq. \ref{eq:pnm_Gtr} where $tr=h$, $G^{h}_{ij}=\qty({1}/{g^{h}_i}+{1}/{g^{h}_{ij}}+{1}/{g^{h}_j})^{-1}$. The hydraulic conductance of pore $i$, $g_i^h$, can be calculated using the Hagen-Poiseuille model \cite{sutera1993} as follows
\begin{equation}
g_i^h=\frac{\pi}{128\mu}\qty(\frac{d_i^4}{l_i}),\label{eq:pnm_gh_3d}
\end{equation}
with $d_i$ being the diameter of pore $i$ and $l_i$ its length. It should be noted here that the length of a pore refers to its radius. The hydraulic conductance of throat $ij$ and pore $j$ are computed in the same manner as for pore $i$. Equation \ref{eq:pnm_gh_3d} is valid for a 3D configuration where the conduit has a cylindrical shape. For a 2D network, where throats are represented by rectangles, the hydraulic conductance is given, from the analytical solution of a plane Poiseuille flow, by
\begin{equation}
g_i^h=\frac{1}{12\mu}\qty(\frac{d_i^3}{l_i}),\label{eq:pnm_gh_2d}
\end{equation}

\subsection{Nernst-Planck Equations}

Special attention was paid to the derivation of the NMEs required to model transport of charged chemical species. In fact, Eq. \ref{eq:np} is discretized in both time and space using various schemes with the resulting accuracy assessed in section \ref{sec:comparisons}. For the sake of brevity in what follows, only semi-discrete forms are presented. First, with a discretized accumulation term (time discretization) and then, with space discretization. One can easily obtain the NME corresponding to Eq. \ref{eq:np} by combining the two semi-discrete forms.

The semi-discrete form of equations \ref{eq:np_con} or \ref{eq:np}, after time discretization, results in the following species $n$ conservation equation
\begin{equation}
\qty[\varphi_b\frac{c^n}{\Delta t}-\varphi_a\qty(-\div{\vb*{N}^{n}})]^{t_1}=\qty[\varphi_b\qty(1-\varphi_a)\qty(-\div{\vb*{N}^{n}})+\varphi_b\frac{c^n}{\Delta t}]^{t_0},\label{eq:np_timediscrete}
\end{equation}
where $\Delta t$ is the time step, $t_0$ the previous time value, $t_1$ the new time value, and $\varphi_a$ and $\varphi_b$ are constants used to set the time scheme. Values $\varphi_a=1$ and $\varphi_b=1$ result in an implicit, first order accurate, time scheme. Whereas setting $\varphi_a=0.5$ and $\varphi_b=1$ corresponds to the second order accurate Crank-Nicolson scheme. Finally, $\varphi_a=1$ and $\varphi_b=0$ yields the steady-state form of the conservation equation.

Focusing on the space discretization of equation \ref{eq:np}, the semi-discrete form can be given by,
\begin{align}
\begin{split}
&\sum_{j=1}^{N_i}\qty[{G}_{ij}^{n,d}+\max\qty(q_{ij}-m_{ij}^{n},0)]c_i^n-\\
&\sum_{j=1}^{N_i}\qty[{G}_{ij}^{n,d}+\max\qty(-q_{ij}+m_{ij}^{n},0)]c_j^n={v}_{i}\pdv{c_{i}^{n}}{t},
\end{split}
& \quad i=1,2,\dots,N_p,\label{eq:pnm_np_1}
\end{align}
such that $c_i^n$ and $c_j^n$ are the concentrations of species $n$ at pore $i$ and neighbor pores $j$, respectively, $G_{ij}^{n,d}$ the diffusive conductance (of species $n$) of the pore-throat-pore assembly, $q_{ij}$ is the throat flow rate, $m_{ij}^{n}$ is the migration rate, and $v_{i}$ the volume of pore $i$. Note that the upwind discretization of the advective and migrative terms should be carried-out considering both terms at the same time as done on Eq. \ref{eq:pnm_np_1}. It was found in this work that considering these terms separately leads to higher errors.

The diffusive conductance $G_{ij}^{n,d}$ of Eq. \ref{eq:pnm_np_1} can be given, based on Eq. \ref{eq:pnm_Gtr}, setting the transport type to $tr=n,d$ to refer to transport of species $n$ via diffusion by $G^{n,d}_{ij}=\qty({1}/{g^{n,d}_i}+{1}/{g^{n,d}_{ij}}+{1}/{g^{n,d}_j})^{-1}$. The pore $i$ diffusive conductance being, for a 3D configuration,
\begin{equation}
g_i^{n,d}=\frac{A_{i}D^{n}}{l_{i}},\label{eq:pnm_g_i}
\end{equation}
such that $A_i$ is the cross-section area of pore $i$, and the diffusion coefficient of species $n$, $D^n$, is considered constant. In the same way as for pore $i$ (Eq. \ref{eq:pnm_g_i}), the diffusive conductances of pore $j$ and throat $ij$ can be defined. For a 2D configuration, the cross-section area $A_{i}$ in Eq. \ref{eq:pnm_g_i} should be replaced by the diameter $d_i$. Furthermore, the volumetric flow rate of the electrolytic solution $q_{ij}$, appearing in Eq. \ref{eq:pnm_np_1}, can be calculated as follows,
\begin{equation}
q_{ij}=G_{ij}^h\qty(p_i-p_j),\label{eq:pnm_q_ij}
\end{equation}
and finally, the migration rate of Eq. \ref{eq:pnm_np_1} can be given under the following form,
\begin{equation}
m_{ij}^{n}=G_{ij}^{n,m}\qty(\phi_i-\phi_j),
\end{equation}
where $G^{n,m}_{ij}=\qty({1}/{g^{n,m}_i}+{1}/{g^{n,m}_{ij}}+{1}/{g^{n,m}_j})^{-1}$ is the migrative conductance and is also defined based on Eq. \ref{eq:pnm_Gtr} where $tr=n,m$ to refer to transport of species $n$ by migration. In these circumstances, the migrative conductance of pore $i$ is
\begin{equation}
g_{i}^{n,m}=\frac{z^{n}F}{RT}g_i^{n,d},
\end{equation}

In equation \ref{eq:pnm_np_1}, while the diffusive flux is discretized based on the central differencing scheme, which is second order accurate in terms of Taylor series expansion, a first order upwind scheme is adopted for both the advective and migration fluxes. However, in a recent work \cite{sadeghi2019a}, a more accurate discretization of the advective and diffusive fluxes was proposed based on the finite difference power-law discretization scheme. Using the power-law discretization for advection and diffusion and the upwind scheme for the migration, the following species conservation equation, that is more accurate than Eq. \ref{eq:pnm_np_1}, can be written
\begin{align}
\begin{split}
&\sum_{j=1}^{N_i}\qty{G_{ij}^{n,d}\max{\qty[\qty(1-\frac{\abs{^{ad}{Pe}^{n}_{ij}}}{10})^{5},0]}+\max{\qty(q_{ij},0)}+\max{\qty(-m_{ij}^{n},0)}}c_i^n-\\
&\sum_{j=1}^{N_i}\qty{G_{ij}^{n,d}\max{\qty[\qty(1-\frac{\abs{^{ad}{Pe}^{n}_{ij}}}{10})^{5},0]}+\max{\qty(-q_{ij},0)}+\max{\qty(m_{ij}^{n},0)}}c_j^n={v}_{i}\pdv{c_{i}^{n}}{t},\\
& \quad i=1,2,\dots,N_p,
\end{split}
\label{eq:pnm_np_2}
\end{align}
where $^{ad}{Pe}^{n}$ is the advective P{\'e}clet number corresponding to species $n$ and is given by the ratio of advective to diffusive contributions as follows
\begin{equation}
^{ad}Pe_{ij}^{n}=\frac{q_{ij}}{G_{ij}^{n,d}}.\label{eq:pnm_pe_1}
\end{equation}

While the discretization given by Eq. \ref{eq:pnm_np_2} is more accurate than Eq. \ref{eq:pnm_np_1}, the migration term, discretized based on an upwind scheme, is only first order accurate and may be a source of non-negligible errors. Indeed, it was shown by \citet{sadeghi2019a}, for advection diffusion problems in pore networks, that the first order upwind discretization of the advective term results in network average relative deviations, in terms of species concentration, of up to $10\%$ compared to FEM simulations. For this reason, an alternative form of the NME was derived where the migration flux was also treated as a power-law. In this form, the advection and migration fluxes in Eq. \ref{eq:np_flux} are grouped together to give rise to a single term that encompasses both advection and migration effects. This leads to an augmented P{\'e}clet number $^{ad,mig}{Pe}^{n}$ which corresponds to the ratio between advective migrative effects and the diffusive ones as follows,
\begin{equation}
^{ad,mig}{Pe}_{ij}^{n}=\frac{q_{ij}-m_{ij}^{n}}{G_{ij}^{n,d}}.\label{eq:pnm_pe_2}
\end{equation}
Accordingly, the species conservation equation takes the following form,

\begin{align}
\begin{split}
&\sum_{j=1}^{N_i}\qty{G_{ij}^{n,d}\max{\qty[\qty(1-\frac{\abs{^{ad,mig}{Pe}^{n}_{ij}}}{10})^{5},0]}+\max{\qty(q_{ij}-m_{ij}^{n},0)}}c_i^n-\\
&\sum_{j=1}^{N_i}\qty{G_{ij}^{n,d}\max{\qty[\qty(1-\frac{\abs{^{ad,mig}{Pe}^{n}_{ij}}}{10})^{5},0]}+\max{\qty(-q_{ij}+m_{ij}^{n},0)}}c_j^n={v}_{i}\pdv{c_{i}^{n}}{t},\\
& \quad i=1,2,\dots,N_p.
\end{split}
\label{eq:pnm_np_3}
\end{align}
For a 2D problem, the volume ${v}_{i}$, appearing in Eqs. \ref{eq:pnm_np_1}, \ref{eq:pnm_np_2}, and \ref{eq:pnm_np_3} has to be replaced by the surface area ${s}_{i}$ to ensure units consistency.

Finally, following the same logic, one can define a migrative P{\'e}clet number which corresponds to the ratio of migrative to diffusive effects,
\begin{equation}
^{mig}{Pe}_{ij}=\frac{-m_{ij}^{n}}{G_{ij}^{n,d}}.\label{eq:pnm_pe_3}
\end{equation}
It can be noticed that $^{mig}{Pe}$, unlike $^{ad}{Pe}^{n}$ and $^{ad,mig}{Pe}^{n}$, does not depend on the chemical species $n$. The $^{mig}{Pe}$ will prove useful in section \ref{sec:comparisons}.

\subsection{Charge Conservation Laws}

As stated above, three different approaches for enforcing charge conservation were considered in this work. The PNM form of each approach is described below. These laws describe the relationship between the electrostatic potential of the solution and the spatial distribution of electric charges in the solution.

\subsubsection{Poisson Equation}

The discretization of the Poisson equation for the electrostatic potential (Eq. \ref{eq:poisson}) is performed based on the second order accurate central differencing scheme. The relative permittivity of the electrolytic solution, $\varepsilon_{r}$, is considered constant and does not depend on the local concentrations. The obtained pore-scale NME, valid for a 3D problem, is given by
\begin{align}
\sum_{j=1}^{N_i}{K^{Poisson}_{ij}}\phi_i-
\sum_{j=1}^{N_i}{K^{Poisson}_{ij}}\phi_j=-{v}_{i}{F}\sum_{n}z^{n}c_{i}^{n},
& \quad i=1,2,\dots,N_p,\label{eq:pnm_poisson}
\end{align}
whereas for a 2D situation, the volume ${v}_{i}$ appearing on Eq. \ref{eq:pnm_poisson} must be replaced by the pore's surface area ${s}_{i}$. In Eq. \ref{eq:pnm_poisson}, $K^{Poisson}_{ij}$ is the ionic conductance of the electrolytic solution for the conduit $ij$. It is given, as on Eq. \ref{eq:pnm_Gtr}, by $K^{Poisson}_{ij}=\qty({1}/{k^{Poisson}_i}+{1}/{k^{Poisson}_{ij}}+{1}/{k^{Poisson}_j})^{-1}$ such that the pore $i$ ionic conductance, for a 3D problem, is
\begin{equation}
{k}^{Poisson}_{i}=\frac{{A}_{i}{\varepsilon\varepsilon_{r}}}{{l}_{i}},\label{eq:pnm_ki_poisson}
\end{equation}
and, for a 2D configuration, it becomes ${k}^{Poisson}_{i}={{d}_{i}{\varepsilon\varepsilon_{r}}}/{{l}_{i}}$. The ionic conductances of pores $j$ neighboring $i$ and the throat $ij$ is computed in the same way as with Eq. \ref{eq:pnm_ki_poisson}.

\subsubsection{Charge Conservation Equation with Electroneutrality}

Charge conservation can also be enforced using Eq. \ref{eq:charge_conservation_02} assuming electroneutrality. The corresponding NME is given as follows
\begin{align}
\begin{split}
&\sum_{j=1}^{N_i}{K}^{elec}_{ij}\phi_i-
\sum_{j=1}^{N_i}{K}^{elec}_{ij}\phi_j=\\
&-F\sum_{n}{z}^{n}\qty(	\sum_{j=1}^{N_i}G_{ij}^{n,d}{c}^{n}_{i}-\sum_{j=1}^{N_i}G_{ij}^{n,d}{c}^{n}_{j}),
\end{split}
& \quad i=1,2,\dots,N_p,\label{eq:pnm_charge_conservation}
\end{align}
where ${K}^{elec}_{ij}$ is the ionic conductance of the electrolytic solution in which electroneutrality is assumed. It is given based on the linear resistor theory for resistors in series (see Eq. \ref{eq:pnm_Gtr}) by $K^{elec}_{ij}=\qty({1}/{k^{elec}_i}+{1}/{k^{elec}_{ij}}+{1}/{k^{elec}_j})^{-1}$ with the ionic conductance for the pore $i$, in a 3D configuration, being,
\begin{equation}
{k}^{elec}_{i}=\frac{{F}^{2}}{RT}\frac{{A}_{i}}{{l}_{i}}\sum_{n}\qty({z}^{n\,2}{D}^{n}{c}_{i}^{n}),\label{eq:pnm_ki}
\end{equation}
and for a 2D problem, ${k}^{elec}_{i}=\qty[{{F}^{2}{d}_{i}}/\qty({RT{l}_{i}})]\sum_{n}\qty({z}^{n\,2}{D}^{n}{c}_{i}^{n})$. Conductances of pores $j$ and throats $ij$ are defined in the same manner as in Eq. \ref{eq:pnm_ki}. For the ionic conductance of throat $ij$, $k^{elec}_{ij}$, the concentration of species $n$ at the considered throat, ${c}^{n}_{ij}$, is required. However, since ${c}^{n}_{ij}$ is not solved for, it can be defined based on a volume (or surface for a 2D problem) weighted average using the concentrations at the two neighbor pores. It is given, for a 3D configuration, by
\begin{equation}
{c}^{n}_{ij}=\frac{{v}_{i}{c}^{n}_{i}+{v}_{j}{c}^{n}_{j}}{{v}_{i}+{v}_{j}},\label{eq:pnm_cij}
\end{equation}
and, becomes ${c}^{n}_{ij}=\qty({{s}_{i}{c}^{n}_{i}+{s}_{j}{c}^{n}_{j}})/\qty({{s}_{i}+{s}_{j}})$, in a 2D problem.

\subsubsection{Laplace Equation}

Finally, another way to enforce charge conservation, is using the Laplace equation for the potential (Eq. \ref{eq:laplace}) in situations where the electrolytic solution is electroneutral and the space variations of the concentration are neglected. The corresponding pore-scale NME is given by,
\begin{align}
\sum_{j=1}^{N_i}{K}^{Laplace}_{ij}\phi_i-\sum_{j=1}^{N_i}{K}^{Laplace}_{ij}\phi_j=0, & \quad i=1,2,\dots,N_p.\label{eq:pnm_laplace}
\end{align}
where ${K}^{Laplace}_{ij}$ is the ionic conductance of the electrolytic solution and is given by $K^{Laplace}_{ij}=\qty({1}/{k^{Laplace}_i}+{1}/{k^{Laplace}_{ij}}+{1}/{k^{Laplace}_j})^{-1}$, in the same manner as for other conductances. The ionic conductance for the pore $i$ is
\begin{equation}
{k}^{Laplace}_{i}=\frac{{A}_{i}}{{l}_{i}},\label{eq:pnm_ki_laplace}
\end{equation}
for a 3D problem, and becomes ${k}^{Laplace}_{i}={{d}_{i}}/{{l}_{i}}$, for a 2D configuration.

\subsection{Solution Algorithm}\label{sec:algorithm}

\begin{figure}
	\centering
	\includegraphics[height=1.25\linewidth]{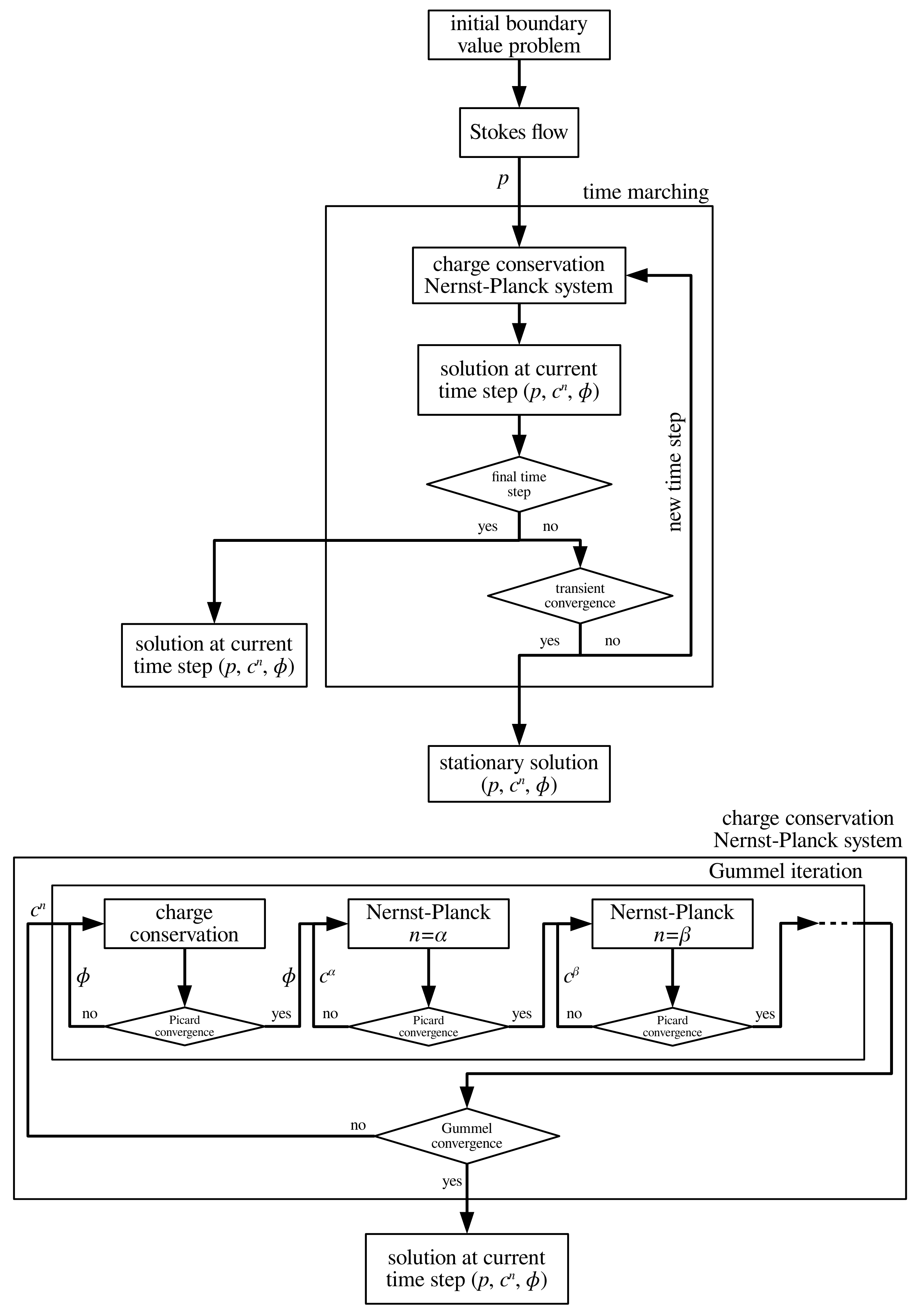}
	\caption{\label{fig:algorithm} Solution algorithm implemented on \texttt{OpenPNM} \cite{gostick2016} to solve time dependent problems of transport of charged chemical species coupled with fluid flow. Fluid flow is described by Eqs. \ref{eq:momentum} and \ref{eq:mass} and the corresponding NME is Eq. \ref{eq:pnm_flow}. A Nernst-Plank equation, Eq. \ref{eq:np} corresponding to NMEs \ref{eq:np_timediscrete} and \ref{eq:pnm_np_1} or \ref{eq:pnm_np_2} or \ref{eq:pnm_np_3}, is adopted for every charged species present in the electrolytic solution. Charge conservation is enforced through Eq. \ref{eq:poisson} or \ref{eq:charge_conservation_02} or \ref{eq:laplace} and the corresponding NMEs are Eqs. \ref{eq:pnm_poisson} or \ref{eq:pnm_charge_conservation} or \ref{eq:pnm_laplace}, respectively.}
\end{figure}
The procedure developed in this work to numerically solve the flow problem (Eqs. \ref{eq:momentum} and \ref{eq:mass}) coupled with the transport of charged species (NP, Eq. \ref{eq:np}, and charge conservation, Eq. \ref{eq:poisson} or \ref{eq:charge_conservation_02} or \ref{eq:laplace} depending on the situation) is described in this section. The solver was implemented within the open-source PNM package \texttt{OpenPNM} \cite{gostick2016}. Although source terms are not considered in sections \ref{sec:background} and \ref{sec:pnm}, the approach followed to handle them is described here. Pore-scale NMEs obtained from the time and space discretization of the PDEs (Eqs. \ref{eq:momentum} and \ref{eq:mass}, Eq. \ref{eq:np}, and Eq. \ref{eq:poisson} or \ref{eq:charge_conservation_02} or \ref{eq:laplace}) are presented in section \ref{sec:pnm}. These NMEs yield linear systems of equations solved iteratively based on the algorithm described on Fig. \ref{fig:algorithm}.

First, the initial and boundary value problem (IBVP), the physical properties of the electrolytic solution, and the solver settings need to be defined. Solver settings include inputs such as the time and space discretization schemes, the different tolerances and maximum number of iterations, type of linear solvers, initial and final time values, the time step, \textit{etc}. Then, the flow problem (Eqs. \ref{eq:momentum} and \ref{eq:mass} corresponding to NME \ref{eq:pnm_flow}) is solved and a converged steady-state pressure field is obtained (see Fig. \ref{fig:algorithm}). Pressure values are used to compute the advective flux in the NP equations.

Subsequently, time marching starts and for each time value, the charge conservation (Eq. \ref{eq:poisson} or \ref{eq:charge_conservation_02} or \ref{eq:laplace} corresponding to NMEs \ref{eq:pnm_poisson} or \ref{eq:pnm_charge_conservation} or \ref{eq:pnm_laplace}, respectively) and NP (Eq. \ref{eq:np} corresponding to NMEs \ref{eq:np_timediscrete} and \ref{eq:pnm_np_1} or \ref{eq:pnm_np_2} or \ref{eq:pnm_np_3}) system is solved based on the Gummel method \cite{jerome1996}. Linear systems are decoupled and solved iteratively and may all be subject to Picard iterations \cite{paniconi1994} in the presence of non-linear source or sink terms. Picard convergence is reached once the value of the solved quantity satisfies the linearized system of equations within a certain tolerance or the maximum number of iterations is reached. The linearization is performed around the value at the previous Picard iteration or the initial value. Gummel iterations are repeated until convergence is obtained or when the maximum number of iterations is reached. A Gummel iteration consists of solving the charge conservation equation, updating the potential values, and solving a NP equation for every species present in the electrolytic solution and finally updating the concentration values. Gummel convergence is achieved when the difference between the values, for both the concentrations and potential, of two successive iterations falls bellow a predefined tolerance. For numerical stability, under-relaxation can be applied to both quantities solved for and/or source or sink terms.

The concentrations and potential fields obtained from the solution of the charge conservation NP system correspond to current time value. The time marching is ended when the predefined final time is reached or if a stationary solution is obtained. Otherwise, a new time iteration will begin after updating all the concentrations and potential values. Stationarity, or transient convergence as shown on Fig. \ref{fig:algorithm}, is obtained once the variation between both concentrations and potential, at two successive time values falls bellow a given tolerance.

\section{Comparisons with Reference Solutions}\label{sec:comparisons}

Ion transport problems over arbitrary disordered porous media were considered here. It is worth recalling that the structural disorder refers to the randomness in pores and throats sizes and in the coordination number of pores. The considered problems were solved numerically based on the PNM approach and, for the sake of comparison, based on the FEM. To assess the accuracy of different NMEs presented in section \ref{sec:pnm}, PNM simulations were performed using three different NMEs. The NMEs consist of Eqs. \ref{eq:pnm_np_1}, \ref{eq:pnm_np_2}, and \ref{eq:pnm_np_3} and are referred to as upwind upwind, power-law upwind, and power-law, respectively. Comparisons focused on the concentration fields only and without losing generality, only one charge conservation scenario was considered for brevity.

\subsection{Initial Boundary Value Problem}\label{sec:IBVP}

\begin{figure}[h]
	\centering
	\includegraphics[width=0.75\linewidth]{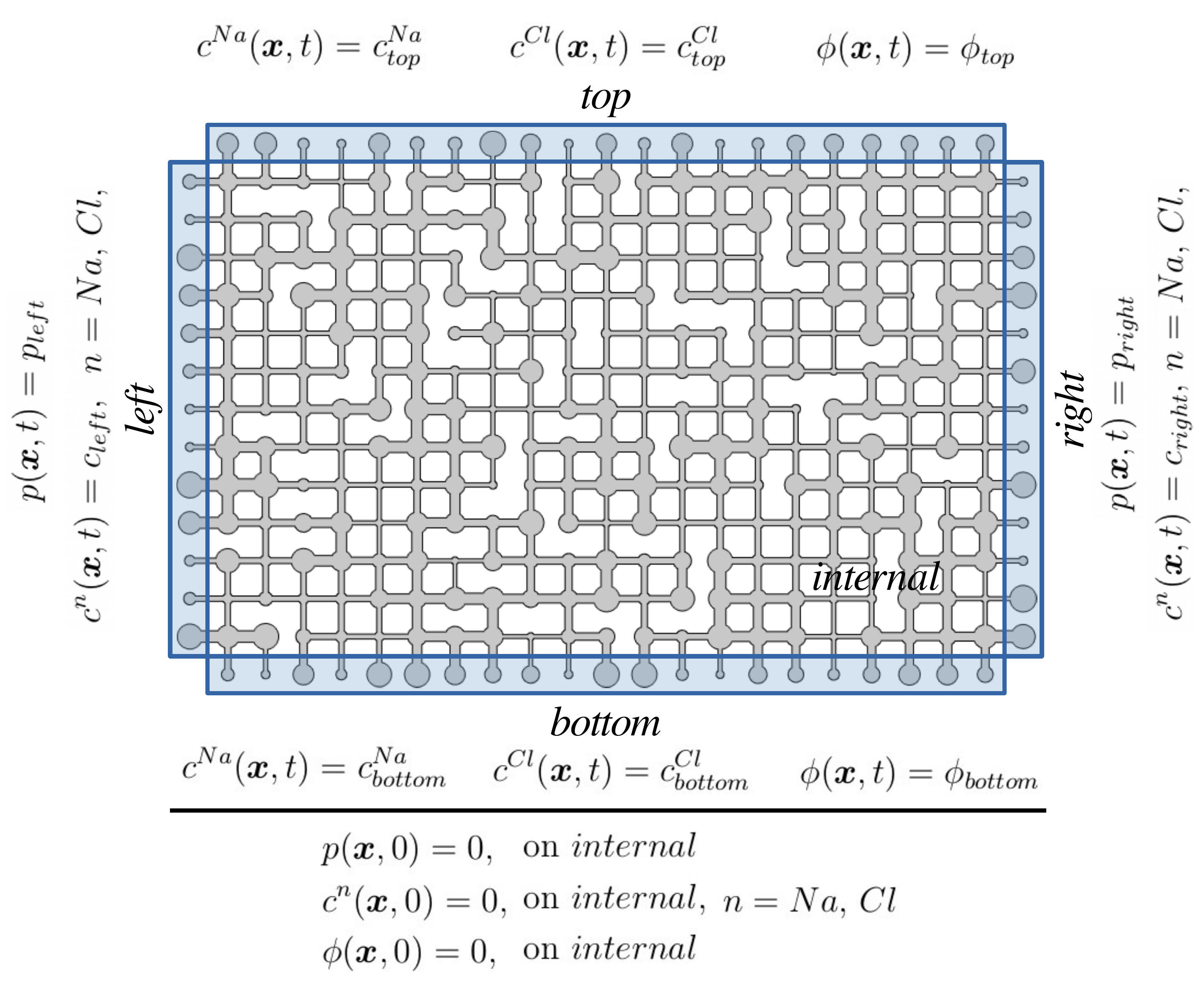}
	\caption{\label{fig:network} A 2D porous realization $\Omega$ made of $341$ pores in a uniform square lattice and connected by throats. Pores and throats have random sizes and spacing between neighbor pores centers is $1\si{\micro m}$. Four boundary regions are defined; $left$, $right$, $bottom$, and $top$, and one internal region; $internal = \Omega \setminus\qty(left\cup right\cup bottom\cup top)$, with the corresponding initial and boundary conditions.}
\end{figure}

The problem under consideration is that of the transport of saline water over an arbitrary porous medium $\Omega$. The real geometry of the 2D porous medium was modeled using a network of pores as shown on Fig. \ref{fig:network}. Despite the fact that the topology of the medium is simplified, analyses based on pore networks were shown to play an important role in diverse applications for the study of flow and transport phenomena in porous media \cite{xiong2016}.

First, a network was generated with $23 \times 15$ pores, connected by throats, consisting of a square lattice with a spacing of $1\si{\micro m}$. Pores and throats were assigned random sizes based on a uniform distribution. The pores at the corners and the throats connecting the boundary pores one to each other were removed for better agreement with the FEM simulations. Finally, the average coordination number of the network was reduced to an average $3$ by deleting random throats not belonging to the minimum spanning tree found using the Kruskal algorithm with random weights assigned to each throat. This increases the structural randomness to more closely mimic real media while remaining geometrically perfectly known. Four boundary regions were defined, namely, $left$, $right$, $bottom$, and $top$, and an internal region $internal = \Omega \setminus\qty(left\cup right\cup bottom\cup top)$. The electrolytic solution (\textit{i.e.}, saline water) is composed of water (solvent) and salt (electrolyte) dissolved and separated into cations, $Na$, and anions, $Cl$. The physical properties of the solution and its components are reported in Tab. \ref{tab:properties}.

\begin{table}
	\centering
	\caption{\label{tab:properties} Physical properties of the mixture (saline water) and its components $Na$ and $Cl$ at temperature $T=298.15~\si{K}$ and pressure $p=101325~\si{Pa}$.}
	\begin{tabular}{lccc}
		\hline
		& mixture & $Na$ & $Cl$ \tabularnewline
		\hline
		Dynamic viscosity ($\mu$) $[\si{Pa\ldotp s}]$ & $0.89557\times10^{-3}$&--&--\tabularnewline
		Relative permittivity ($\varepsilon_{r}$) & $78.303$&--&--\tabularnewline
		Diffusivity ($D^{n}$) $[\si{m^{2}/s}]$ & -- & $1.33\times10^{-9}$ &$2.03\times10^{-9}$\tabularnewline
		Valence ($z^{n}$) & -- & $+1$ &$-1$\tabularnewline
		\hline
	\end{tabular}
\end{table}

The flow of the mixture is described by Eqs. \ref{eq:momentum} and \ref{eq:mass} whereas the movement of ions is modeled using Eq. \ref{eq:np} and the charge conservation is enforced through Eq. \ref{eq:laplace}. The initial and boundary conditions associated with this system of equations are included in Fig. \ref{fig:network}. Boundary concentrations are $c_{left}=10\si{mol/m^3}$, $c_{right}=20\si{mol/m^3}$, $c^{Na}_{bottom}=c^{Cl}_{top}=5\si{mol/m^3}$, and $c^{Na}_{top}=c^{Cl}_{bottom}=30\si{mol/m^3}$. Although the considered transport problem is arbitrary and is only used for comparisons between different methods, the configuration is comparable to what occurs in a spacer of a desalination unit by capacitive deionization \cite{hemmatifar2015}. The analysis is performed in terms of the network's arithmetic mean of the absolute values of the dimensionless numbers $^{ad}{Pe}^{Na}$ and $^{mig}{Pe}$ referred to as $\langle ^{ad}{Pe}^{Na} \rangle$ and $\langle ^{mig}{Pe} \rangle$, respectively. These numbers were varied by considering different values for the pairs $p_{left}$ $p_{right}$ and $\phi_{bottom}$ $\phi_{top}$, respectively. The considered simulation conditions are such that both $\langle ^{ad}{Pe}^{Na} \rangle$ and $\langle ^{mig}{Pe} \rangle$ were varied within a range from $0.1$ to $5$ considering all possible combinations. The network-scale advective forces were always kept acting from $right$ to $left$ by enforcing $p_{right}>p_{left}$. On the other hand, migration influences the transport of ions in a perpendicular direction depending on the ions valence. For $Na$, migration occurs from $bottom$ to $top$ since $\phi_{bottom}>\phi_{top}$.

\subsection{Numerical Considerations}\label{sec:numerical}

The transport problems were solved numerically based on the PNM approach described on section \ref{sec:pnm} using \texttt{OpenPNM} \cite{gostick2016}. The FEM simulations were performed using \texttt{COMSOL} \cite{comsol2018}.

\begin{figure}
	\centering
	\includegraphics[width=0.7\linewidth]{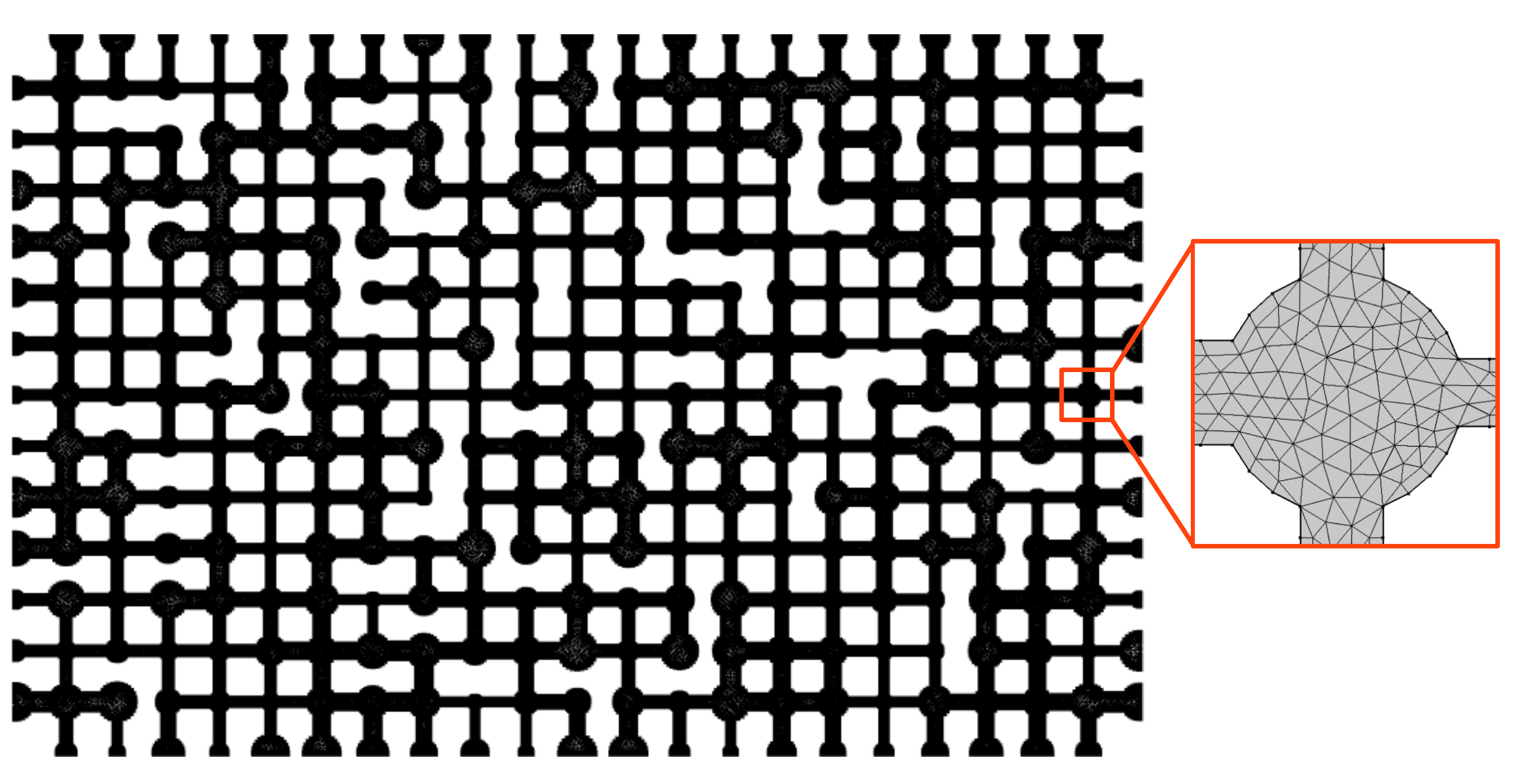}
	\caption{\label{fig:mesh} Computational domain modeling the geometry of Fig. \ref{fig:network} with the corresponding grid used for FEM simulations.}
\end{figure}

For FEM simulations, the boundary pores defined on Fig. \ref{fig:network} were trimmed at the plan passing through their centers as shown on Fig. \ref{fig:mesh}. The boundary conditions are then imposed on the resulting boundary edges. This approach is adopted in order to impose comparable simulation conditions on both the PNM and FEM simulations since boundary conditions are imposed at the pore centers in the PNM simulations. For FEM simulations, the computational domain was meshed, after a mesh sensitivity analysis, using a grid comprised of $97598$ elements for a medium including 341 pores. Triangular and quadrilateral elements were used (see Fig. \ref{fig:mesh}).

The FEM simulations were performed using the \texttt{Creeping Flow}, \texttt{Transport of Diluted Species}, and \texttt{Laplace Equation} modules. The system of non-linear equations was solved using Newton's method and at each of its iterations, the linearized system was solved using the multifrontal massively parallel sparse direct solver MUMPS \cite{amestoy2006}. For consistency, the same linear solver was used with the PNM simulations.

\subsection{Simulation Results}\label{sec:results}

\begin{figure}[h]
	\centering
	\includegraphics[height=0.67\linewidth]{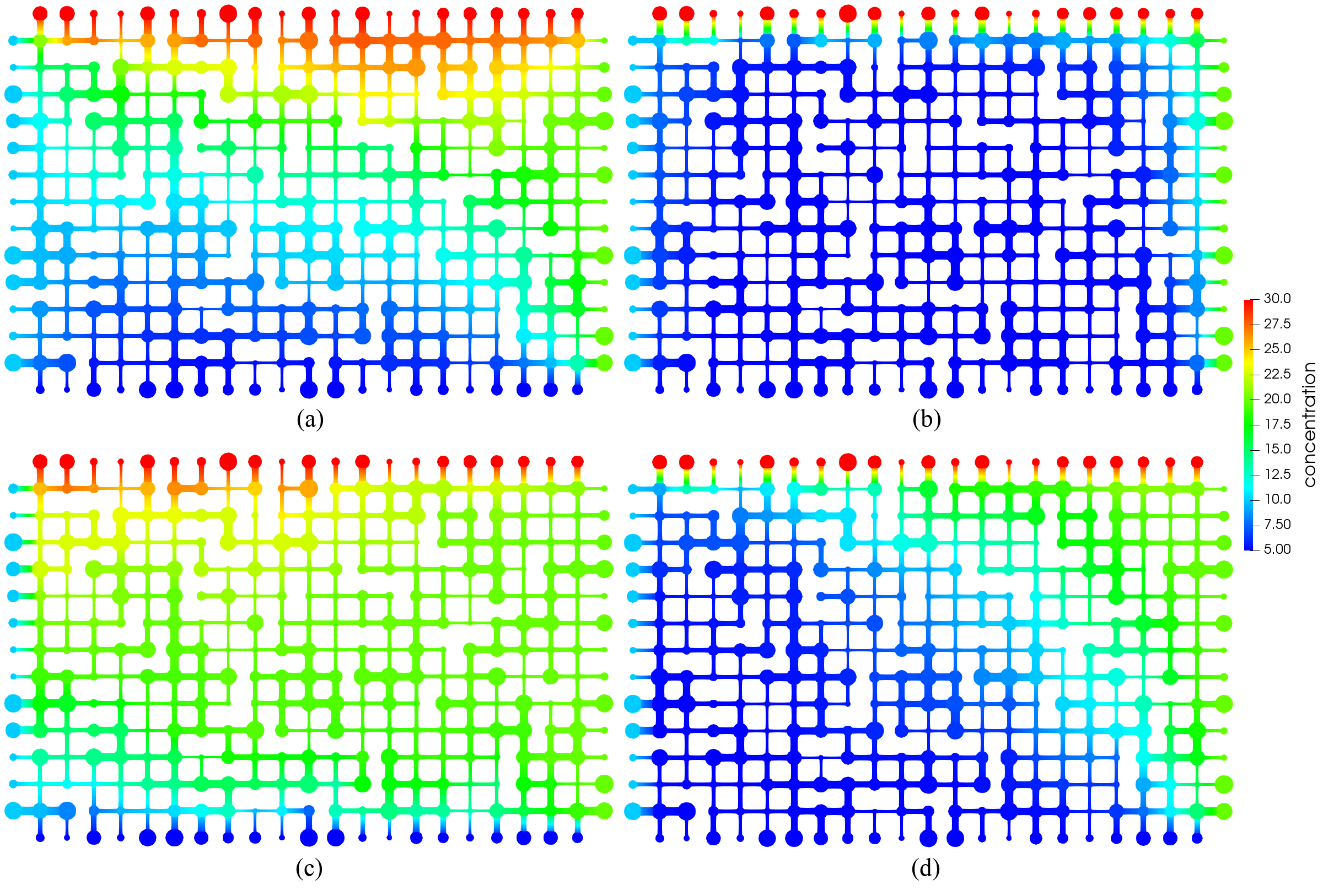}
	\caption{\label{fig:concentration} Concentration of $Na$ color map at steady state obtained from PNM simulations based on the power-law NME (Eq. \ref{eq:pnm_np_3}). Values at the throats are obtained from the interpolation of the neighbor pores concentrations. Simulation conditions: (a) $\langle ^{ad}{Pe}^{Na} \rangle=0.1$, $\langle ^{mig}{Pe} \rangle=0.1$, (b) $\langle ^{ad}{Pe}^{Na} \rangle=0.1$, $\langle ^{mig}{Pe} \rangle=5$, (c) $\langle ^{ad}{Pe}^{Na} \rangle=5$, $\langle ^{mig}{Pe} \rangle=0.1$, and (d) $\langle ^{ad}{Pe}^{Na} \rangle=5$, $\langle ^{mig}{Pe} \rangle=5$.}
\end{figure}

Figure \ref{fig:concentration} shows the $Na$ concentration color map obtained from the solution of the problem defined above (section \ref{sec:IBVP}) for some of the considered configurations. These results were obtained based on the PNM approach using the power-law NME (Eq. \ref{eq:pnm_np_3}). This figure shows that, for the considered problems, when advection and migration forces act with comparable intensities at the network scale, more heterogeneous $Na$ concentration distributions are obtained (Figs. \ref{fig:concentration} (a) and (d)). This is due to the fact that boundaries over which these two forces are imposed are at different uniform concentrations. When one of the these two transport mechanisms dominates, a more uniform concentration field is observed (Figs. \ref{fig:concentration} (b) and (c)) since uniform concentration values are imposed at the boundaries.

The solutions obtained from the FEM simulations are not shown on Fig. \ref{fig:concentration} as they are comparable to the PNM ones with a negligible deviation discussed below. The deviation between PNM and FEM simulations, in terms of concentration of species $n$ at the center of pore $i$ at steady state, is given by
\begin{align}
E^n_i=\frac{\abs{c^n_{i,PNM}-c^n_{i,FEM}}}{c^n_{i,FEM}}, & & \quad n=Na,\,Cl, & & \quad i=1,2,\dots,N_p,\label{eq:error}
\end{align}
where the FEM solution is considered as the reference one. In Eq. \ref{eq:error}, $c^n_{i,PNM}$ and $c^n_{i,FEM}$ are concentrations of species $n$ at the center of pore $i$ obtained from PNM and FEM simulations, respectively. The analysis of the deviation was carried-out based on the arithmetic mean of $\abs{E^{Na}_i}$ over the entire network and is referred to as $\sigma$.

\begin{figure}
	\centering
	\includegraphics[height=1.24\linewidth]{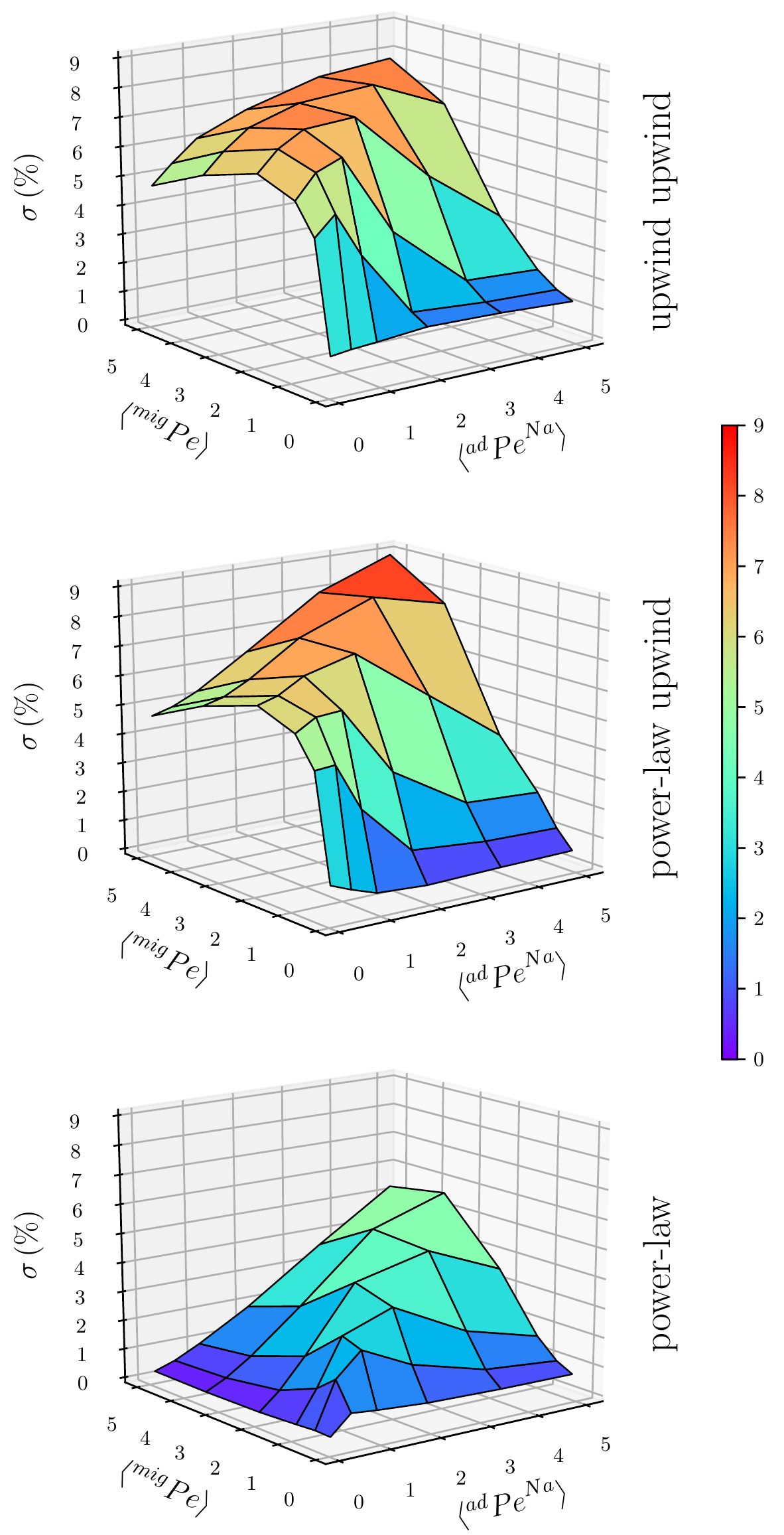}
	\caption{\label{fig:deviations} Color map of $\sigma$ versus advective $\langle ^{ad}{Pe}^{Na} \rangle$ and migrative $\langle ^{mig}{Pe} \rangle$ P\'eclet numbers at steady state. $\sigma$ is the arithmetic mean of the absolute deviation between $Na$ concentrations obtained from PNM and FEM simulations (see Eq. \ref{eq:error}). PNM simulations based on the upwind upwind (Eq. \ref{eq:pnm_np_1}), power-law upwind (Eq. \ref{eq:pnm_np_2}), and power-law (Eq. \ref{eq:pnm_np_3}) NMEs. Initial and boundary value problem defined in section \ref{sec:IBVP}.}
\end{figure}

Values of $\sigma$ obtained using the upwind upwind, power-law upwind, and power-law NMEs are shown on Fig. \ref{fig:deviations}. Although the deviation $\sigma$ is always below an acceptable value of $9\%$, local deviations of up to $50\%$ were observed with the two former NMEs for certain configurations. This is consistent with a recent work \cite{sadeghi2019a} where large deviations between PNM and FEM were observed on dispersion problems in pore networks when the upwind scheme was used in PNM simulations. It was also reported that, for certain advection diffusion problems, the deviation between PNM and FEM simulations increases with the advective P\'eclet number \cite{sadeghi2019a}. This is also seen in the results reported on Fig. \ref{fig:deviations}. The difference in the dependence of $\sigma$ on advective and migrative P\'eclet numbers at low values can be attributed to the transport configuration adopted here where the advective, diffusive and migrative driving forces act in different directions in the network.

Analysis of Fig. \ref{fig:deviations} also shows that similar behaviors are obtained with the upwind upwind and power-law upwind NMEs although the latter globally presents slightly lower deviations. On the other hand, a significant decrease in $\sigma$ is obtained with the power-law NME. In fact the average deviation is consistently below $5\%$ and marginally exceeds this value when $\langle ^{ad}{Pe}^{Na} \rangle=5$ and $\langle ^{mig}{Pe} \rangle \geq 3.5$. For migration diffusion dominated transport ($\langle ^{ad}{Pe}^{Na} \rangle \leq 0.1$), which is of practical relevance for applications such as battery simulations, a negligible (below $0.4\%$) deviation is observed. The same applies when transport is advection diffusion dominated ($\langle ^{mig}{Pe} \rangle \leq 0.1$), which is of importance for dispersion problems, where $\sigma \leq 1$. It can be concluded from this analysis that the power-law NME should be used when performing PNM simulations to ensure a maximum accuracy.

\begin{figure}
	\centering
	\includegraphics[height=0.67\linewidth]{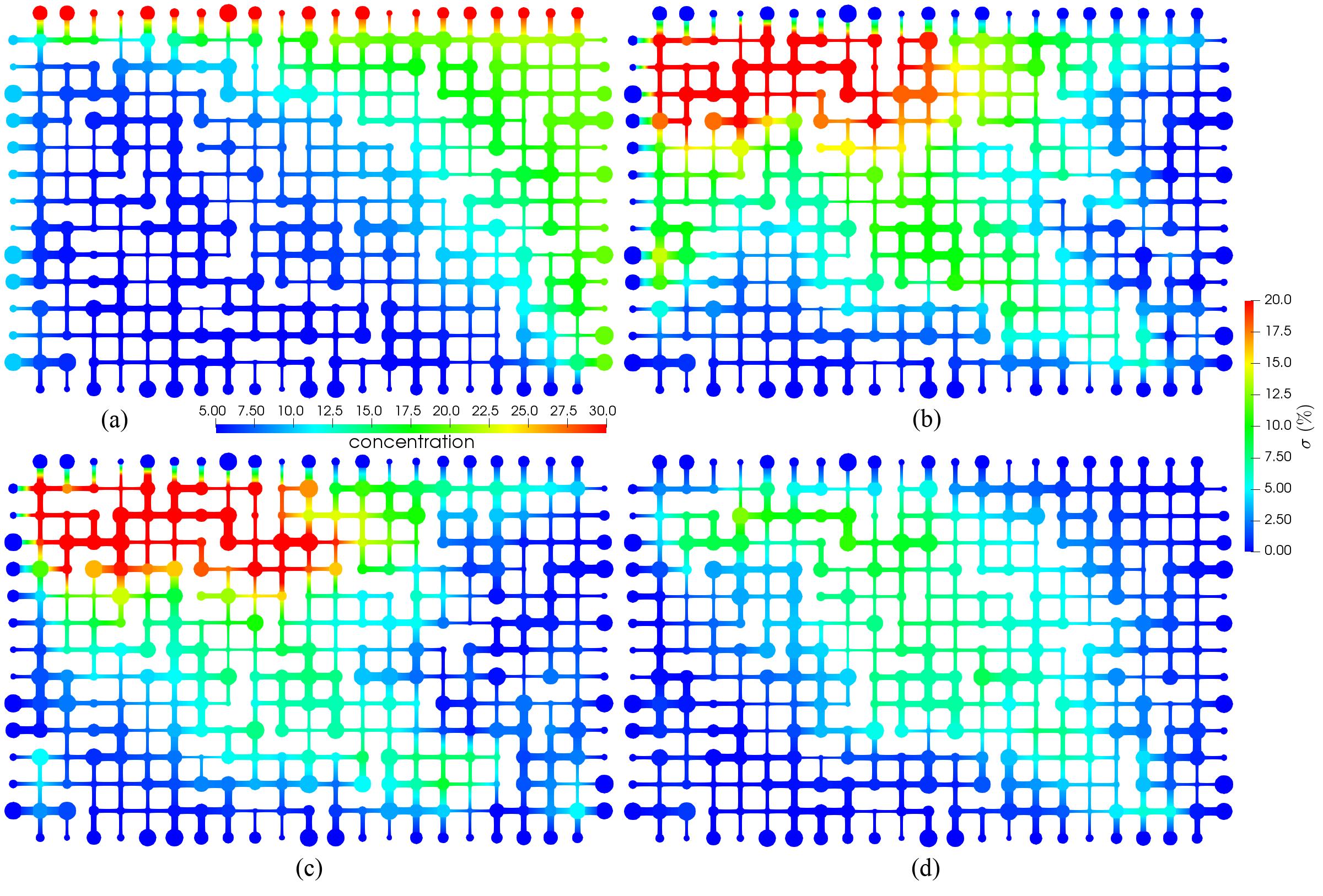}
	\caption{\label{fig:concentration_dev} (a) Concentration of $Na$ color map at steady state obtained from FEM simulations. Color map of the deviation between PNM and FEM simulations $\sigma$ (see Eq. \ref{eq:error}) such that PNM simulations are based on the (b) upwind upwind (Eq. \ref{eq:pnm_np_1}), (c) power-law upwind (Eq. \ref{eq:pnm_np_2}), and (d) power-law (Eq. \ref{eq:pnm_np_3}) NMEs. Simulation conditions: $\langle ^{ad}{Pe}^{Na} \rangle=1$, $\langle ^{mig}{Pe} \rangle=1$. Initial and boundary value problem defined in section \ref{sec:IBVP}.}
\end{figure}

The source of the deviations between the PNM and FEM simulations resulting from the use of the upwind scheme were discussed in detail in a recent work \cite{sadeghi2019a}. They were attributed to the fact that in the presence of moderate to important advective effects (\textit{i.e.}, P\'eclet numbers equal or larger than unity), significant local concentration gradients appear, and the assumption of linear concentration profiles between pores loses accuracy. This behavior also appears in Fig. \ref{fig:concentration_dev}. The considered transport configuration gives rise to a high concentration front on the diagonal of the porous medium from the upper left to the bottom right vertices (see Fig. \ref{fig:concentration_dev} (a)). The high deviation regions coincide with this front for the different NMEs (Figs. \ref{fig:concentration_dev} (b), (c), and (d)).

\begin{figure}
	\centering
	\includegraphics[height=0.37\linewidth]{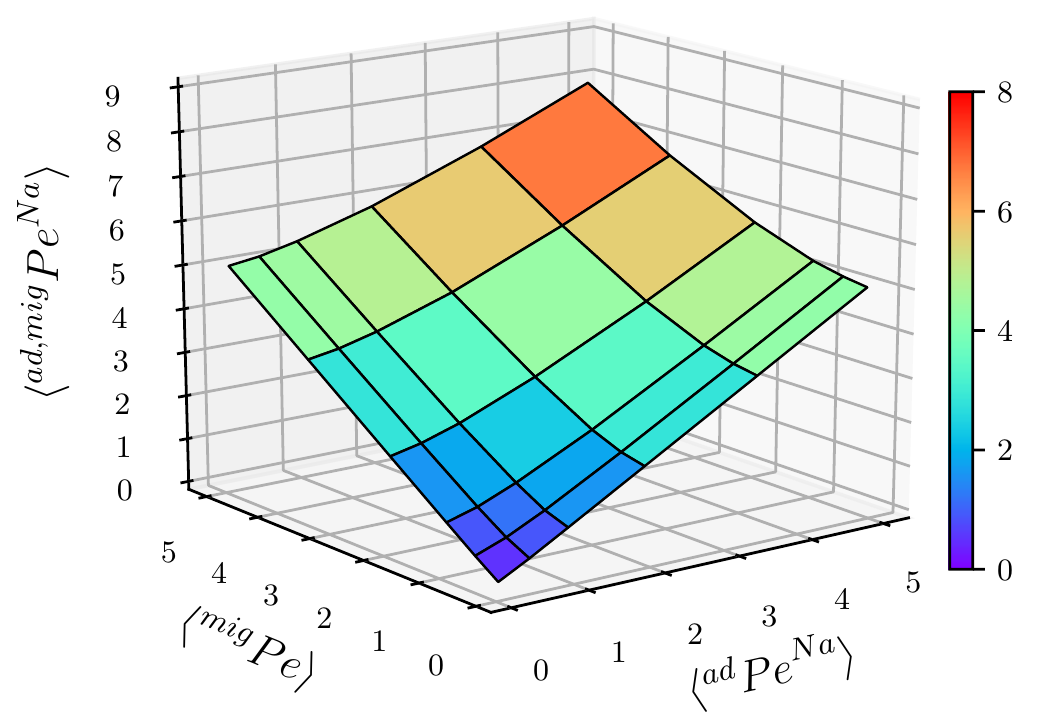}
	\caption{\label{fig:peclet} Network scale augmented P\'eclet number $\langle ^{ad,mig}{Pe}^{Na} \rangle$ versus the advective $\langle ^{ad}{Pe}^{Na} \rangle$ and migrative $\langle ^{mig}{Pe} \rangle$ ones. P\'eclet numbers obtained from the network's arithmetic mean of the absolute value of pore-scale P\'eclet numbers given by Eqs. \ref{eq:pnm_pe_1}, \ref{eq:pnm_pe_2} and \ref{eq:pnm_pe_3}. Initial and boundary value problem defined in section \ref{sec:IBVP}.}
\end{figure}

Finally, the conclusions drawn from the analysis of Fig. \ref{fig:deviations}, based on $\langle ^{ad}{Pe}^{Na} \rangle$ and $\langle ^{mig}{Pe} \rangle$, can be generalized to be valid when one considers $\langle ^{ad,mig}{Pe}^{Na} \rangle$. In fact, from Fig. \ref{fig:peclet}, for the considered problems, $\langle ^{ad,mig}{Pe}^{Na} \rangle$ has a quasi-linear dependence upon $\langle ^{ad}{Pe}^{Na} \rangle$ and $\langle ^{mig}{Pe} \rangle$. Hence, the deviation between PNM and FEM increases with $\langle ^{ad,mig}{Pe}^{Na} \rangle$.

\subsection{Simulation Time}\label{sec:time}

\begin{figure}
	\centering
	\includegraphics[height=0.4\linewidth]{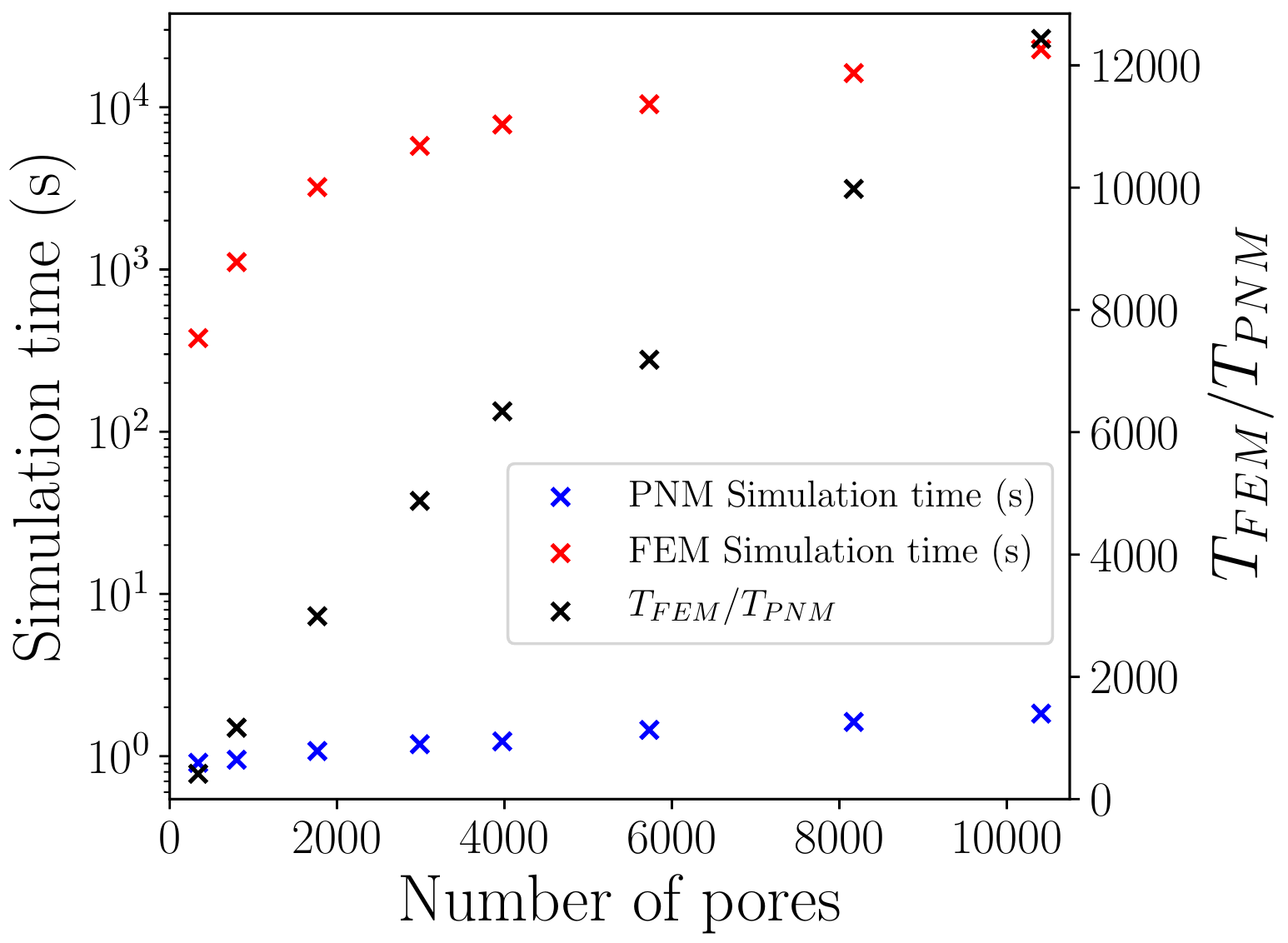}
	\caption{\label{fig:times_and_performance} Simulation time using the PNM (\texttt{OpenPNM} \cite{gostick2016}) and FEM (\texttt{COMSOL} \cite{comsol2018}) solvers and their ratio versus the size of the porous medium (\textit{i.e.}, number of pores). Network scale advective and migrative P\'eclet numbers set to $\langle ^{ad}{Pe}^{Na} \rangle=1$ and $\langle ^{mig}{Pe} \rangle=1$, respectively. Initial and boundary value problem defined in section \ref{sec:IBVP}. Simulations run in parallel using two X5650 Intel Xeon CPUs at $2.67\si{GHz}$ with $12$ cores in total.}
\end{figure}

The reduced computational cost of the PNM approach over FEM is staggering. The size of the medium was characterized considering the number of pores included while following the same approach described in section \ref{sec:IBVP} to generate the domains. Simulations were run on two X5650 Intel Xeon CPUs at $2.67\si{GHz}$ with $12$ cores in total. The meshing time on the FEM simulations is not included in the comparisons for consistency, although it also requires important computational resources. In fact, meshing the largest domain (includes $10410$ pores), performed in parallel on $12$ cores, took $1219\si{s}$ for a total of $\sim3.04\times{10}^{6}$ grid cells. Whereas generating a cubic network, even with millions of pores is almost instantaneous.  

Figure \ref{fig:times_and_performance} shows the simulation time versus the number of pores, $Np$, for PNM and FEM approaches. For the $Np$ range investigated here, both approaches show a quasi-linear dependence upon $Np$. The simulation time scales as $T_{PNM}(\si{s})=9.08\times{10}^{-5}Np+0.89$ and $T_{FEM}(\si{s})=2.05Np$ with the PNM and FEM solvers considered in the present work, respectively. This means that for the considered range of network sizes, the simulation time increases more than $22.5\times{10}^{3}$ times faster with the FEM solver compared to the PNM one. For the largest computational domain analyzed here, comprising $10410$ pores, solution of the transport problem was performed in just $1.83\si{s}$ using \texttt{OpenPNM}. On the other hand, $\sim3.4\si{h}$ were needed for the FEM simulation using \texttt{COMSOL}. This result highlights the significant decrease in simulation time which can be achieved adopting the PNM approach described in section \ref{sec:algorithm}, even for the coupled non-linear multiphysics problem studied here.

The ratio between simulation times using the PNM (\texttt{OpenPNM} \cite{gostick2016}) and FEM (\texttt{COMSOL} \cite{comsol2018}) solvers $T_{FEM}/T_{PNM}$ versus the size of the porous medium is also reported on Fig. \ref{fig:times_and_performance}. It can be seen that simulation speedup increases with the number of pores reaching a speedup factor of over ${10}^{4}$ for a medium including $\sim{10}^{4}$ pores. The speedup is expected to increase for the same number of pores when considering 3D porous media. In addition to the simulation speedup obtained with the PNM approach compared to the FEM one, the PNM simulations can be run using limited memory resources. In this study, carrying-out the FEM simulation on the largest domain considered (comprising $10410$ pores) required $\sim96\si{GB}$ of memory while only $241.4\si{MB}$ were used on the PNM simulation.

\section{Conclusions}\label{sec:conclusions}

Ion transport problems in pore networks with random pore sizes and coordination numbers were considered and solved numerically using PNM and FEM solvers. The transport was modeled based on the NP equations for each charged species present in the electrolytic solution in addition to a charge conservation equation which relates the concentration of different species one to each other. In the presence of a fluid flow, the momentum and mass conservation equations, were adopted to describe the fluid flow.

Several time and space discretization schemes were presented to derive the NMEs corresponding to the considered PDEs. The accuracy of each scheme was compared to a reference solution generated by FEM, and best agreement was found when a power-law approach was applied to both the advection diffusion and migration terms. This is consistent with our previous work on advection diffusion \cite{sadeghi2019a}. These model equations were implemented within the open-source package \texttt{OpenPNM} \cite{gostick2016} based on the Gummel algorithm with relaxation. Comparisons showed a maximum relative deviation, in terms of ions concentration, between PNM and FEM below $\sim5\%$ with the PNM simulations being over ${10}^{4}$ times faster than the FEM ones on a medium including ${10}^{4}$ 2D pores. The speedup is expected to increase for the same number of pores when considering 3D porous media.

The PNM approach allows for simulations with significantly lower computational costs compared to other DNS methods, while retaining reasonable accuracy. This will allow for more effective design and analysis or operation for many electrochemical systems since computation can be performed on large samples while retaining pore-scale resolution. Ultimately, this highly-efficient computational framework could be used for optimization of electrode architectures and cell designs \cite{fornercuenca2019}. 

\section*{Computer Code Availability}

The developed solver for transport of charged species in porous media is available on \texttt{OpenPNM} \cite{gostick2016} public repository \textcolor{blue}{\href{https://github.com/PMEAL/OpenPNM}{https://github.com/PMEAL/OpenPNM}}.

\section*{Acknowledgments}

This research was supported by CANARIE Canada.

\bibliographystyle{elsarticle-harv}
\bibliography{references}
\end{document}